\documentclass[12pt]{article}
\usepackage{epsfig}
  
\def\be{\begin{equation}}
\def\ee{\end{equation}}
\def\ben{\begin{displaymath}}
\def\een{\end{displaymath}}
\def\ba{\begin{array}{c}}
\def\bal{\begin{array}{l}}
\def\ea{\end{array}}

\begin{document}

 \begin{center}
{\tiny .}

\vspace{.35cm}

 {\Large \bf
 Conditional
  observability
   }\end{center}

\vspace{10mm}

 \begin{center}

 {\bf Miloslav Znojil}

 \vspace{3mm}
Nuclear Physics Institute ASCR,

 250 68 \v{R}e\v{z}, Czech Republic

{e-mail: znojil@ujf.cas.cz}

\vspace{3mm}

\vspace{5mm}
%


\end{center}

\vspace{5mm}

\section*{Abstract}
For a quantum Hamiltonian $H=H(\lambda)$, the observability of the
energies $E$ may be robust (whenever all $E$ are real at all
$\lambda$) or, otherwise, conditional. Using a pseudo-Hermitian
family of $N-$state chain models $H=H^{(N)}(\lambda)$ we discuss
some generic properties of conditionally observable spectra.

\newpage

\section{\label{par1} Introduction}

At the low energies and for the sufficiently weak interactions,
quantum mechanics is a reliable theory. A transition to the
full-fledged apparatus of relativistic quantum field theory is,
moreover, generally believed to offer its natural extension to the
higher energies and/or to the stronger interaction forces. In
between these two extremes, unfortunately, there exists a huge
territory of open theoretical questions. The loss of the internal
consistency of many approximative phenomenological models may be
encountered. Their limits of validity at certain parameters are
often marked by the loss of the reality of the measurable
quantities. One of the best known illustrations of such a
``quantum catastrophe" is the complexification of the ground-state
energy in the solvable model of a Dirac fermion moving in a
overcritical external Coulomb field \cite{Fluegge}. The same
effect is encountered for the Klein-Gordon boson moving in an
external scalar field \cite{MZ}, etc.

A particularly popular simulation of the breakdown of quantum
stability has recently been found via the study of complex, ${\cal
PT}-$symmetric local potentials admitting the completely real
spectra \cite{BB}. Such an extension of the usual phenomenology
offers multiple advantages in the context of physics (cf., e.g.,
its recent review \cite{Carl}). At the same time, its proper
implementation requires a fairly complicated mathematical
apparatus \cite{DDT}. This is a serious technical obstacle for any
easy explanation of the parameter-controlled transitions between
the real and complex energies. In fact, within the class of the
${\cal PT}-$symmetric local potentials the reality of the spectrum
may prove extremely fragile \cite{MZ2}. For this reason one feels
inclined to turn attention to the constructive study of the
control of the observability of the energies (i.e., of the reality
of {\em all} the eigenvalues of $H$) via some simplified models
specified, say, by some finite-dimensional matrix Hamiltonians.

The first results in this direction were already obtained in our
two recent remarks \cite{PLBh,PLBB} where we discussed some
phenomenological aspects of certain ``first nontrivial", viz.,
two- and three-dimensional real-matrix models, respectively. In
our present continuation of this effort we intend to extend our
attention to the whole family of simplified matrix models of an
arbitrary (say, even) dimension $N=2J$.

The key encouragement of such a project has been found in our
computer-algebra-based paper~\cite{maximal} where we revealed that
a transparent mathematical structure can emerge not only in the
above-mentioned low-dimensional models $H^{(2)}$ and $H^{(3)}$ but
also in some of their specific tridiagonal generalizations
$H^{(N)}$ of {\em any} dimension $N$. For our present purposes we
pick up a subset of the latter models with the even dimensions
$N=2J$. The reason is pragmatic -- one finds just purely formal
differences between the separate even- and odd-dimensional series
of the models~$H^{(N)}$ of ref.~\cite{maximal}.

Our main message will be preceded by a brief summary of the state
of the art in section \ref{prveky}. A more detailed explanation of
the problem will then follow in section~\ref{prvak}. We emphasize
there that several relevant properties of our family of the
Hamiltonians may already be observed in its first, one-parametric
real-matrix two-by-two member $H^{(2)}(\lambda)$. We point out the
intimate connection between the parity-pseudo-Hermiticity of the
model and the mechanism which makes the ``observability" (i.e.,
the reality) of the energies lost during a smooth change of the
$\lambda-$dependent matrix elements.

The next steps of our detailed analysis of chain-models
$H^{(N)}(\lambda)$ are performed in sections \ref{druhak} and
\ref{tretak} giving a detailed description of the spectra at $N=4$
and $N=6$, respectively. In section \ref{generel} some of the
quantitative conclusions of this analysis are found extensible to
all the dimensions $N=2J$. In particular, within the
``catastrophic" loss-of-the-observability scenario we stress that
the flexibility offered by the $J$ independent matrix elements in
$H^{(2J)}$ {\em suffices} for the simulation and enumeration of
the eligible level-confluences igniting the imminent quantum
collapse.

Section \ref{katak} is a summary where we re-emphasize an
exceptional suitability of our present models for deductions of
some universal and generic qualitative conjectures.

\section{The loss of observability in
${\cal PT}-$symmetric examples \label{prveky} }

A deep physical appeal of the ${\cal PT}-$symmetric quantum
mechanics (PTSQM, \cite{BBjmp}) lies in the variability of its
{\em definition} of the inner product in the ``physical" Hilbert
space of states \cite{Geyer,alirev,BBJ}. The idea itself is in
fact not too surprising since it has been discovered, forgotten
and rediscovered in field theory \cite{Nagy}, in
perturbation-theory mathematics \cite{Caliceti,BG} as well as in
nuclear physics etc \cite{Geyer,Hatano}. After its recent
extremely successful popularization by Bender and Boettcher
\cite{BB} its use helped to clarify also the stability of a
quantum particle in many {\em complex} quantum potentials
\cite{mytri} -- \cite{all}.

A mathematical core of the PTSQM formalism lies in its work with
pseudo-Hermitian Hamiltonians $H$ such that $H^\dagger = {\cal
P}\,H\,{\cal P}^{-1}\neq H$. The intertwiner ${\cal P}$ should be
an elementary operator which is, very often, identified with the
parity \cite{BB,BG}. Then, the manifestly non-Hermitian
Hamiltonian $H$ can only become selfadjoint (or, in the language
of ref. \cite{Geyer}, quasi-Hermitian) with respect to a
nontrivial, {\it ad hoc} metric operator $\Theta \neq I$ which
depends on $H=H(\lambda)$ and, hence, which can also vary with the
parameter $\lambda$.

The selected $\Theta$ defines the inner product in the Hilbert
space of states ${\cal H}^{(physical)}$ so that it may become
singular at some ``exceptional" $\lambda$ \cite{Langer}. This is
precisely what paved the way to the practical physical
applications of the conditional reality of the eigenvalues in the
single-particle relativistic context of Klein-Gordon equation
\cite{Ali} and of Proca equation \cite{Smejkal} as well as in the
papers on quantum anomalies \cite{Jones}, on supersymmetric models
\cite{Bagchi} or, beyond quantum-theory context, in cosmology
\cite{WdW} and in magnetohydrodynamics \cite{Oleg}.

We believe that in virtually all of these (and many other)
applications the conditional character of the reality of the
eigenvalues plays the key role. For this reason let us now perform
a certain systematic clarification of the related phenomenological
as well as purely formal possibilities.

\begin{figure}[t]                     
\begin{center}                         
\epsfig{file=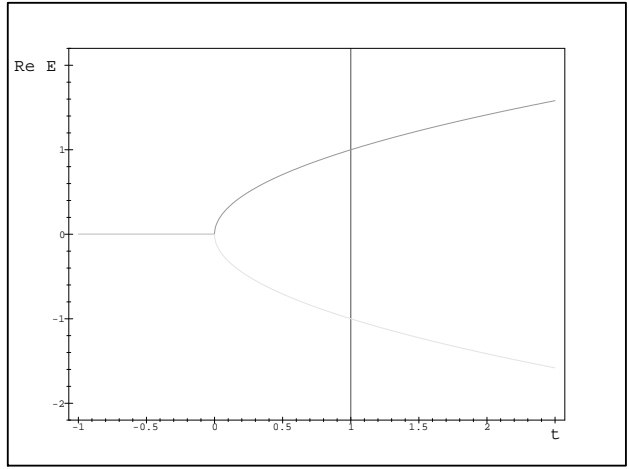,angle=270,width=0.6\textwidth}
\end{center}                         
\vspace{-2mm} \caption{Real parts of the energies in the two-state
model as functions of the parameter $t$.
 \label{obr1a}}
\end{figure}

\begin{figure}[t]                     
\begin{center}                         
\epsfig{file=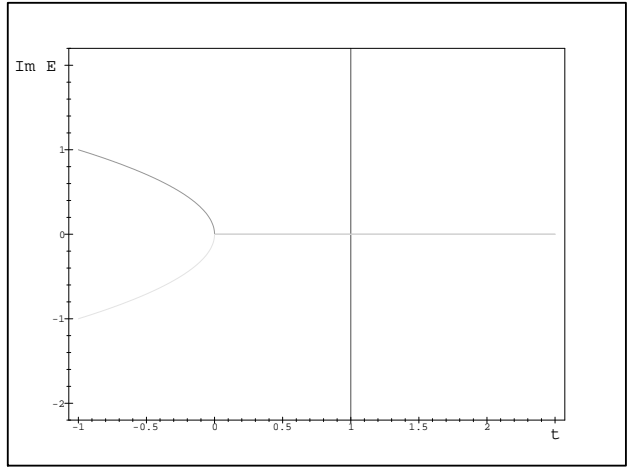,angle=270,width=0.6\textwidth}
\end{center}                         
\vspace{-2mm} \caption{Imaginary parts of the energies in the
two-state model as functions of the parameter $t$.
 \label{obr1b}}
\end{figure}

\section{The loss/emergence of observability
 \label{prvak}
in a matrix example}

\begin{figure}[t]                     
\begin{center}                         
\epsfig{file=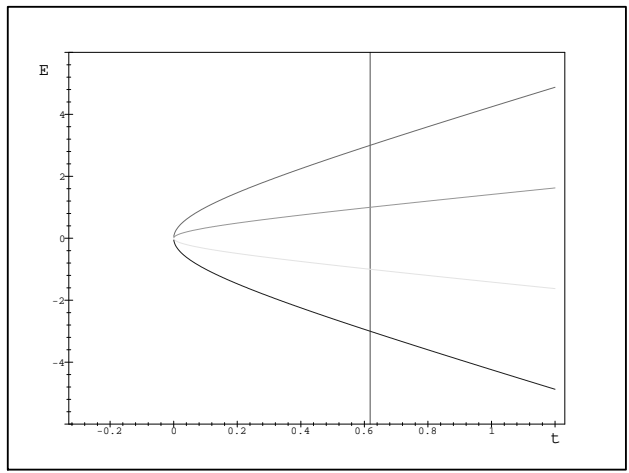,angle=270,width=0.6\textwidth}
\end{center}                         
\vspace{-2mm} \caption{The four real $J=2$ energies at $A=B=1$
 \label{obr2a}}
\end{figure}

\begin{figure}[t]                     
\begin{center}                         
\epsfig{file=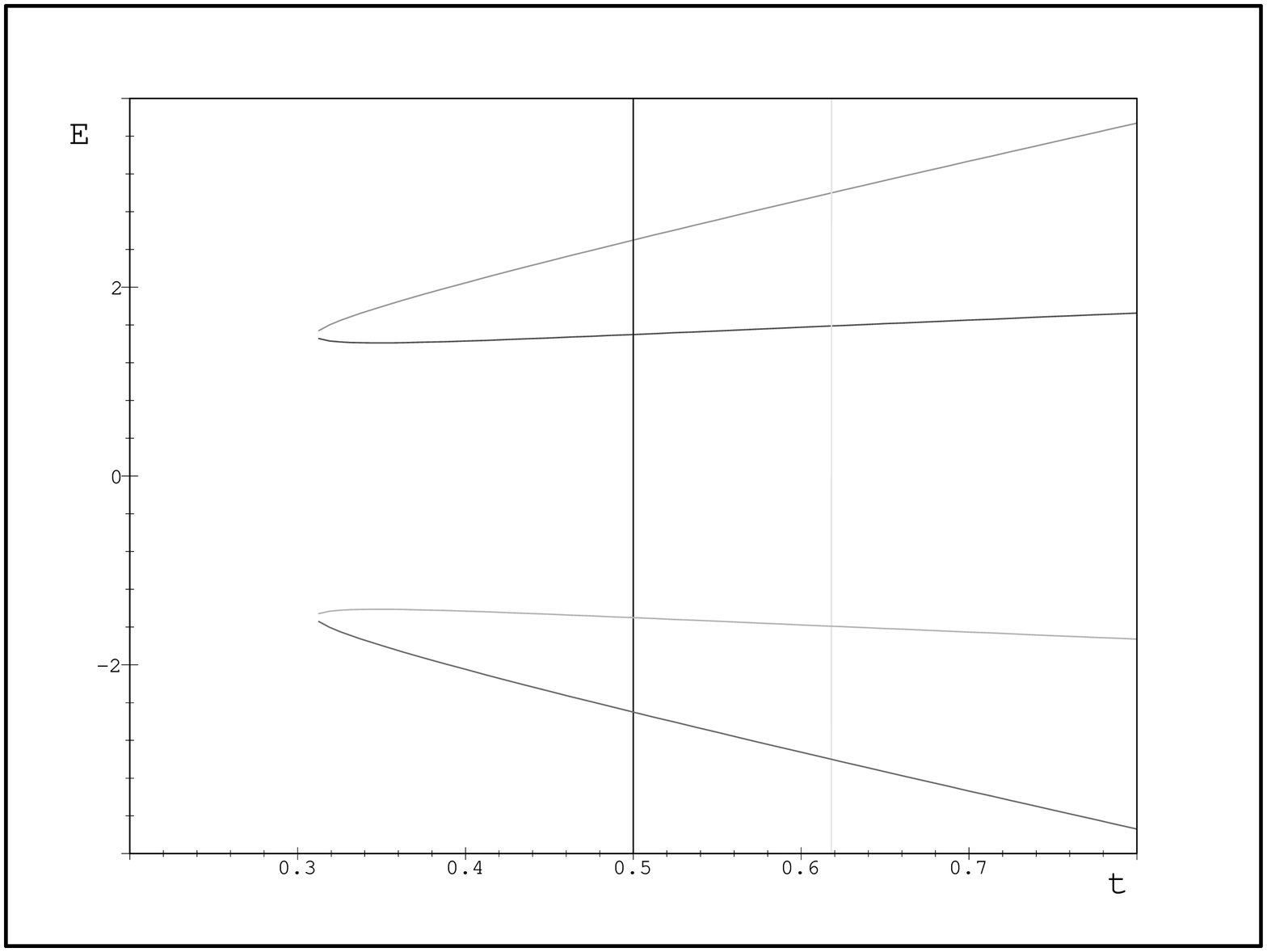,angle=270,width=0.6\textwidth}
\end{center}                         
\vspace{-2mm} \caption{The four real $J=2$ energies at $A/2=B=1$
 \label{obr2b}}
\end{figure}

The feasibility of a controlled, conditional transition to the
instability characterized by the complex eigenvalues may be
understood as a direct consequence of the non-Hermitian nature of
the models $H=H(\lambda)$ for which the parameter $\lambda$ either
stays inside its ``allowed" physical domain ${\cal D}$ or some of
the energy eigenvalues $E$ become complex. Of course, the
variations of $\lambda$ may be expected to depend on time. For
this reason, let us change our notation and replace $\lambda$ by
the more explicit symbol $t$.

Whenever our ``time" $t$ reaches its critical value, the related
complexification transition may be most easily visualized in the
schematic two-by-two model of ref.~\cite{PLBh},
 \be
 H^{(2)}= \left (
 \begin{array}{cc}
 -1&\sqrt{1-t}\\
 -\sqrt{1-t}&1
 \ea
 \right )\,.
 \label{dvehl}
 \ee
At all $t < 1$ this matrix is real and pseudo-Hermitian with
respect to certain parity matrix ${\cal P}$,
 \be
 \left [ H^{(2)}\right ]^\dagger = {\cal P}
 H^{(2)}
  {\cal P}^{-1}\,,
  \ \ \ \ \ \  \ \ \ \ {\cal P}
  = \left (
 \begin{array}{cc}
 1&0\\
 0&-1
 \ea
 \right )\,.
 \label{dvepahl}
 \ee
The specific parametrization of its matrix elements has been
chosen as giving closed formula for the two-point spectrum,
$E_\pm^{(2)} = \pm \sqrt{t}$. Obviously, these energies generated
by the Hamiltonian (\ref{dvehl}) remain complex (i.e., not
observable) along all the negative half-axis, $t \in (-\infty,0)$
(cf. Figures \ref{obr1a} and \ref{obr1b}). On the contrary, the
matrix $H^{(2)}(t)$ becomes manifestly Hermitian (we could say,
``conventionally physical") at $t \in (1,\infty)$. In the middle,
``unconventionally physical" interval of parameters $t \in (0,1)$,
our non-Hermitian Hamiltonian with real energies remains, in the
terminology of the review paper \cite{Geyer}, {\em
quasi-Hermitian}.

The ``unconventional" choice of $t \in (0,1)$ leaves our matrix
$H^{(2)}$ tractable as a valid and acceptable selfadjoint
representation of an observable quantity, provided only that our
Hilbert space of states ${\cal H}^{(2)}$ is properly re-defined
(cf. the extensive accounts of this attitude, say, in the reviews
\cite{Carl,alirev} or in the proceedings \cite{proc}). In this
sense the message delivered by our illustrative example
(\ref{dvehl}) may be read as emphasizing that even though its form
may cease to be {\em manifestly} Hermitian, the operator $H$ may
represent a physical observable (comment \cite{PLBh} may be
consulted for all the technical details).

Even though the symbol $t$ represents time, let us keep {\em both}
the {directions} of the $t-$development available. Thus, under the
tacit assumption that we move to the left along the real $t-$axis
in Figures (\ref{obr1a}) and (\ref{obr1b}) we speak about the
confluence of the real energy levels followed by their subsequent
complexification. This convention will be preferred in what
follows although, alternatively, we could also change it and call
the right-ward $t-$development a ``big-bang-like"
decomplexification of the system which stayed unobservable at the
earlier times.

Our specific $t-$parametrization of the matrix elements is
privileged because it mediates the smooth $t-$dependence of the
eigenvalues (or of their squares at least). Thus, even when the
formal Hermiticity and/or quasi-Hermiticity of the Hamiltonian
$H^{(2)}$ breaks down, both the Figures \ref{obr1a} and
\ref{obr1b} confirm that the $t=1$ division line between the
Hermitian and non-Hermitian regime is artificial and, in any
conceivable phenomenological context, irrelevant.

Our two latter comments are challenging: it is not obvious what
can be expected to happen at some higher dimensions in a suitable
matrix generalization of $H^{(2)}$. A few answers will be offered
in what follows.

\begin{figure}[t]                     
\begin{center}                         
\epsfig{file=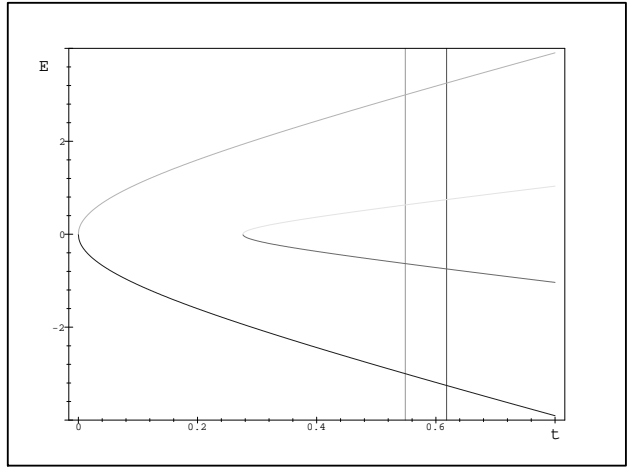,angle=270,width=0.6\textwidth}
\end{center}                         
\vspace{-2mm} \caption{The four real $J=2$ energies at $A=2B/3=1$
 \label{obr2c}}
\end{figure}

\begin{figure}[t]                     
\begin{center}                         
\epsfig{file=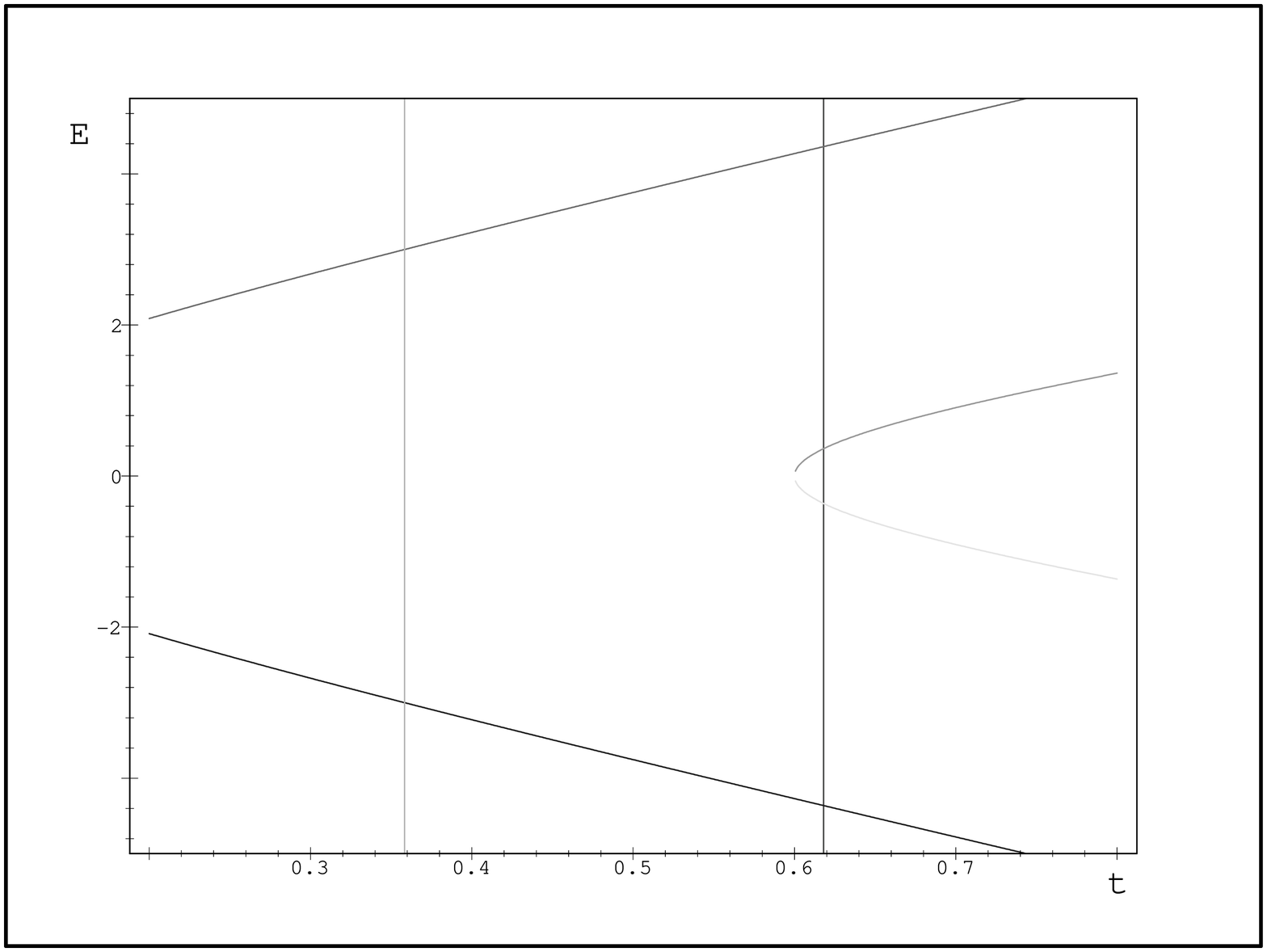,angle=270,width=0.6\textwidth}
\end{center}                         
\vspace{-2mm} \caption{The four real $J=2$ energies at $A=B/5=1$
 \label{obr2d}}
\end{figure}

\section{The first nontrivial model \label{druhak} with four levels }

The two-parametric four-by-four chain-model
 \be
 H^{(4)}=
 \left (\begin {array}{cccc}
  -3&\sqrt {3-3\,\beta}&0&0\\
  -\sqrt {3-3\,\beta}&-1&
 2\,\sqrt {1-\alpha}&0\\
 0&-2\,\sqrt {1-\alpha}&1&\sqrt {3-3\,\beta}\\
 0&0&-\sqrt {3-3\,\beta}&3
 \end {array}
 \right )\,
 \label{ham4}
  \ee
is just one of the special cases of the four-parametric
pseudo-Hermitian Hamiltonians studied in ref.~\cite{PLA}. The
quadruplet of the eigenenergies of matrix (\ref{ham4}) is
obtainable in closed form,
  \be
  E_\pm=\pm\sqrt{s}\,,
  \ \ \ \ \ \
 s=s_\pm=3\,\beta+2\,\alpha \pm 2\,\sqrt {3\,\beta\,
 \alpha+{\alpha}^{2}+9\,\beta-9\,\alpha}\,
 \label{ruty}
  \ee
so that the boundary of the two-dimensional domain ${\cal D}$
(where our pseudo-Hermitian matrix has the real spectrum) is
composed of the two curves, viz.,
 \be
 \beta\geq \beta_{minimal}=\frac{9\,\alpha-{\alpha}^{2}}
 {9+3\,
 \alpha}\,, \ \ \ \ \ \ \ \ \ \ \ \
 \alpha\in (0,1)
 \,
 \label{uchyla}
 \ee
and
 \be
  \alpha\geq \alpha_{minimal} = \beta-\frac{\beta^2}{4}\,,
   \ \ \ \ \ \ \ \ \ \ \ \
 \beta\in (0,1)
 \,.
 \label{uchylbe}
  \ee
Once we set
 \be
 \beta=t+B\,t^2\,,\ \ \ \ \ \ \alpha=t+A\,t^2\,,
 \label{ans4}
 \ee
we may fix the auxiliary constants $A=B=1$ and stay safely inside
the quasi-Hermiticity domain ${\cal D}$ at all the sufficiently
small values of $t>0$. In a way illustrated by Figure \ref{obr2a}
the $t-$dependence of the energies remains smooth also when we
cross the separation point $t=t^{(GM)}=(\sqrt{5}-1)/2\approx
0.618$ between the Hermitian and non-Hermitian regimes with
$t>t^{(GM)}$ and $t<t^{(GM)}$, respectively.

Once we decide to weaken the attraction between the central
energies and choose, say, $A=2$ and $B=1$, the energies split in
the two well-separated pairs and they remain all real whenever
$t^{(QH)}>0.3104686356$, i.e., to the right from the $A-$dependent
quasi-Hermiticity boundary. The resulting $t-$dependence of the
spectrum is displayed in Figure  \ref{obr2b} where the
complexifications of the two different off-diagonal matrix
elements are marked by the two different vertical lines. To the
right of both of them, our Hamiltonian $H^{(4)}$ becomes standard
and Hermitian, to the left of both of them, our matrix $H^{(4)}$
remains ${\cal P}-$pseudo-Hermitian with respect to the usual
parity
 \be
 {\cal P}=
 \left (\begin {array}{cccc}
  1&0&0&0\\
 0&-1&
 0&0\\
 0&0&1&0\\
 0&0&0&-1
 \end {array}
 \right )\,.
 \label{standa}
  \ee
Between the vertical lines, an anomalous parity matrix must be
chosen to define the pseudo-Hermiticity,
 \ben
 {\cal P}'=
 \left (\begin {array}{cccc}
  -1&0&0&0\\
 0&1&
 0&0\\
 0&0&1&0\\
 0&0&0&-1
 \end {array}
 \right )\,.
  \een
In the complementary scenario we have to weaken the attraction
between the peripheral energy levels using $B>A$. The choice of
$A=1$ accompanied by the weakly enhanced $B=3/2$ leads to the
result depicted in Figure  \ref{obr2c}. Due to the
complexification of the most strongly attracted central pair of
the levels, the quasi-Hermiticity is lost for
$t<t^{(QH)}=0.2761423749$. This takes place safely below the upper
boundary $t^{(SPH)}=0.5485837704$ of the pseudo-Hermitian regime
specified by the standard parity operator (\ref{standa}). The
latter feature is fragile. At the larger $B=5$ we get
$t^{(SPH)}=0.3582575695$ which is perceivably smaller than the
complexification bound $t^{(QH)}=3/5$ (cf. Figure \ref{obr2d}).

\begin{figure}[t]                     
\begin{center}                         
\epsfig{file=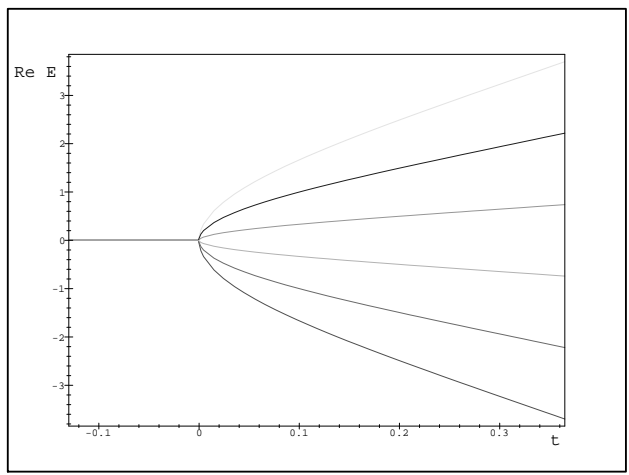,angle=270,width=0.6\textwidth}
\end{center}                         
\vspace{-2mm} \caption{Real parts of the energies in the six-state
model ($A=B=C=1$).
 \label{obr3a}}
\end{figure}

\begin{figure}[t]                     
\begin{center}                         
\epsfig{file=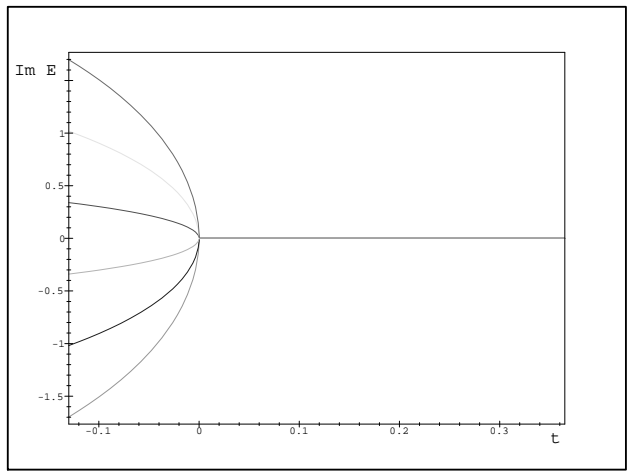,angle=270,width=0.6\textwidth}
\end{center}                         
\vspace{-2mm} \caption{Imaginary parts of the energies in the
six-state model ($A=B=C=1$).
 \label{obr3b}}
\end{figure}

\section{The next, more complicated chain model with six levels
 \label{tretak} }

\begin{figure}[t]                     
\begin{center}                         
\epsfig{file=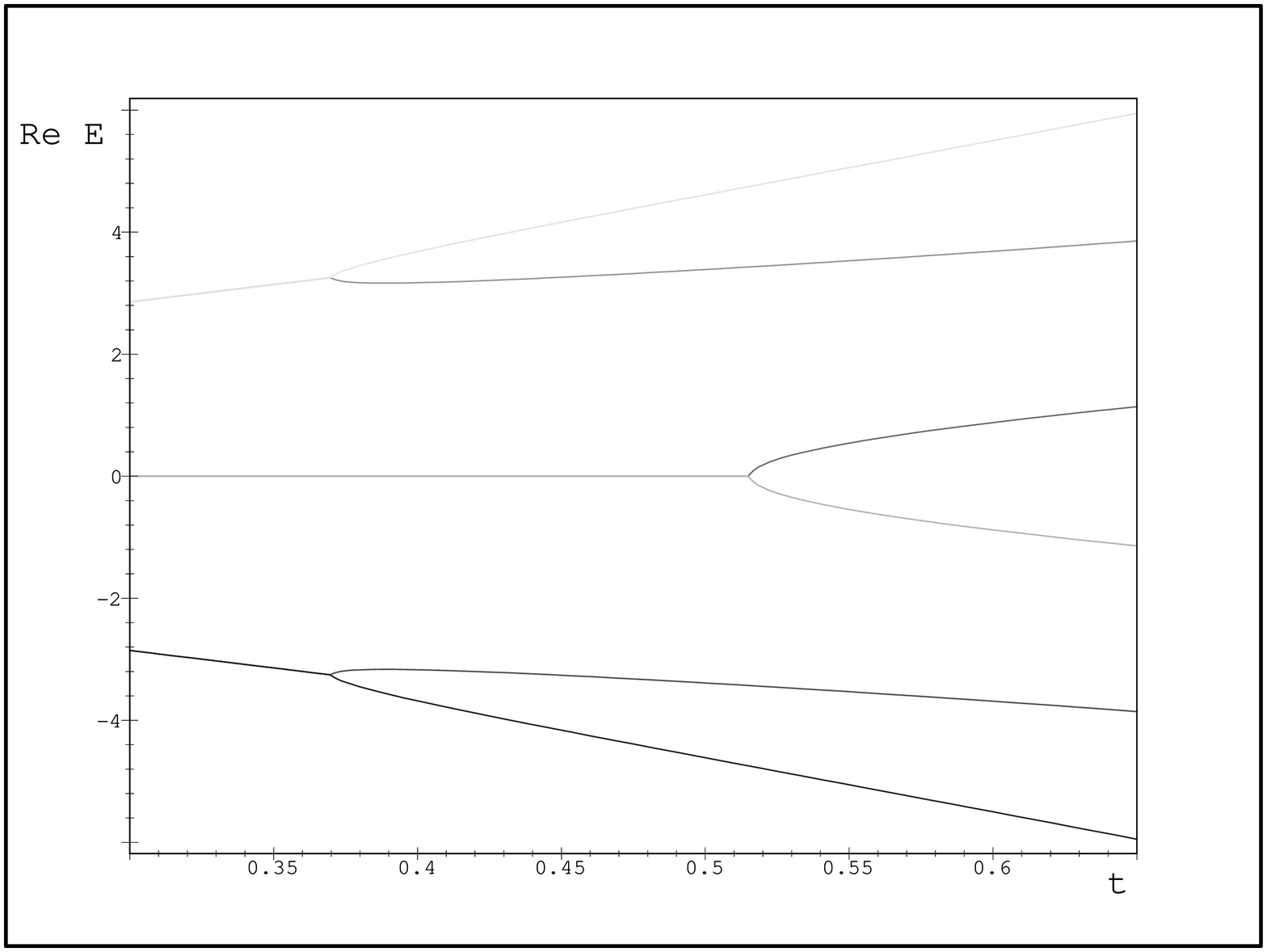,angle=270,width=0.6\textwidth}
\end{center}                         
\vspace{-2mm} \caption{Real parts of the energies in the six-state
model at $A=B/2=C=1$.
 \label{obr4}}
\end{figure}

\begin{figure}[t]                     
\begin{center}                         
\epsfig{file=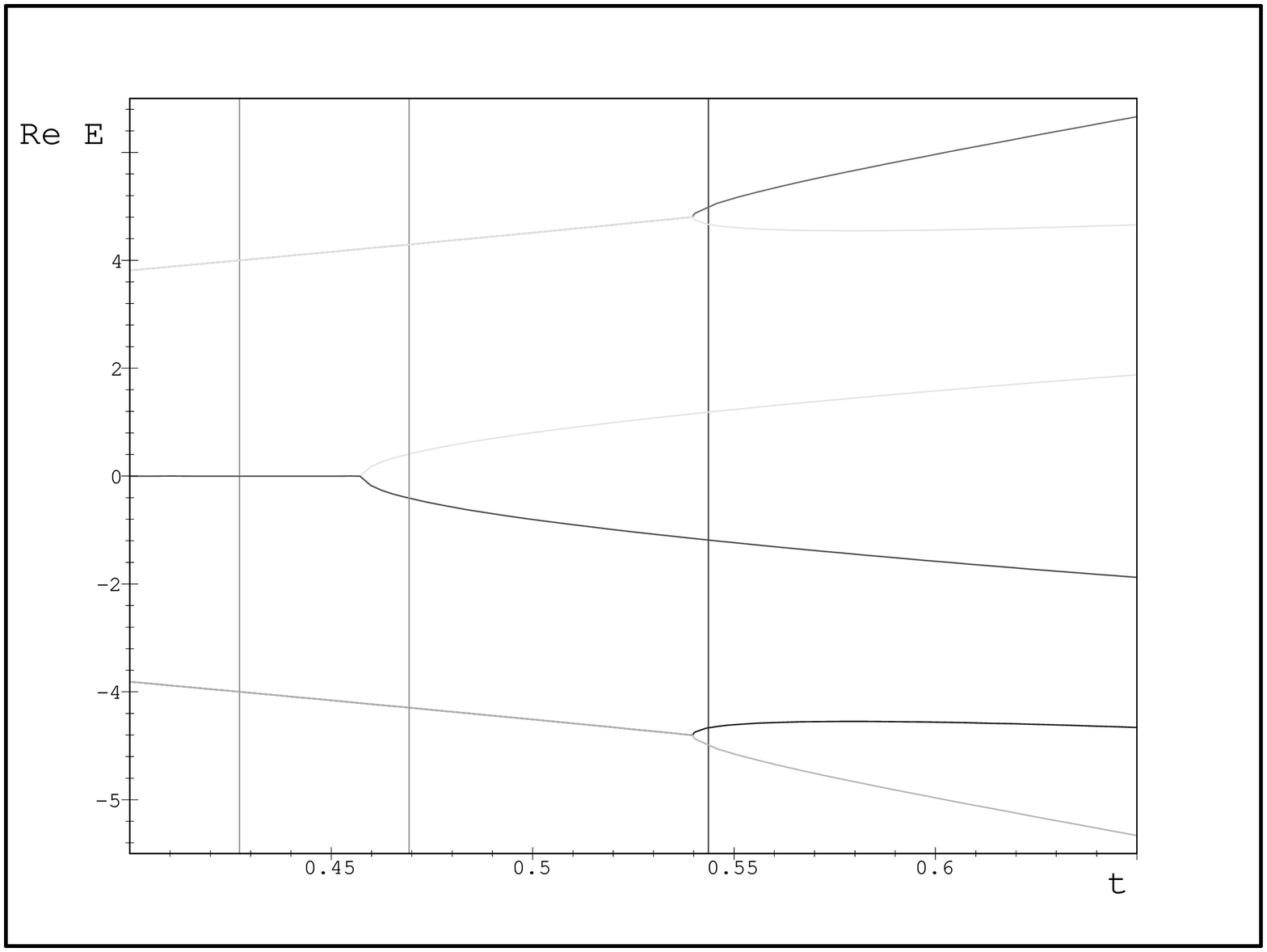,angle=270,width=0.6\textwidth}
\end{center}                         
\vspace{-2mm} \caption{Real parts of the energies in the six-state
model at $A/3=B/5=C=1$.
 \label{obr5}}
\end{figure}

In the six-by-six model of ref. \cite{maximal},
  \be
 H^{(6)}
 =\left [\begin {array}{cccccc}
  -5&g_1&0&0&0&0\\
 -g_1& -3&g_{2}&0&0&0\\
 0&-g_{2}&-1&g_3&0&0
 \\
 0&0&-g_{3}&1&g_{2}&0
 \\
 0&0&0&-g_2&3&g_{1}\\
 0&0&0&0&-g_{1}&5
 \end {array}\right ]\,,
 \label{hamm}
 \label{6IPTS}
 \ee
with
 \ben
 g_1=c=\sqrt{5\,(1-\gamma)}\,,\ \ \ g_2=b= 2\,\sqrt{2\,(1-\beta)}\,,
 \ \ \ g_3=a= 3\,\sqrt{1-\alpha}
 \label{sol3}
 \een
let us set
 \be
  \alpha = t+t^2+A\,t^3 \,,
 \ \ \ \ \
  \beta = t+t^2+B\,t^3\,,
 \ \ \ \ \
  \gamma = t+t^2+C\,t^3\,.
 \label{ans6}
  \label{malik}
 \ee
For $A=B=C=1$ we encounter a not too interesting single
pseudo-Hermiticity line $t^{(PH)}=0.5436890127$ which is,
incidentally, out of the range of Figures \ref{obr3a} and
\ref{obr3b}. Thus, under our present interpretation of $t$ as a
leftwards-running time, the choice of $A=B=C=1$ would represent a
fine-tuned ``big crunch" process. Let us add a comment that after
a temporary reversal of our conventional arrow of time, Figure
\ref{obr3a} mimics an even more interesting scenario of a
``big-bang" development. From this point of view there exists no
observable state  of our schematic system at $t<0$ and  just a
single, fully degenerate state with $E=0$ emerges at $t=0$.
Subsequently, teh evolution becomes characterized by a steady
repulsion of the levels.

After we return to our leftwards-running time convention, the
peripheral-weakening choice of a dominant $C$ will leave both the
outermost levels far away, too weakly attracted and not
sufficiently participating in the overall collapsing tendency. The
first complexification will involve only the inner quadruplet of
the energies.  It may proceed {\em either}  along the
two-pair-complexification pattern indicated in Figure \ref{obr2b}
(we choose there $A=B=C/2=1$ so that the complementary
pseudo-Hermiticity boundary moved to $t^{(PH')}=1/2$) {\em or}
along the central-complexification pattern represented by Figures
\ref{obr2c} and  \ref{obr2d} (necessitating an increase of  $B$).

The next eligible choice of dominant $A$ will weaken the central
attraction so that the overall loss of the quasi-Hermiticity will
be caused by the simultaneous pairwise mergers inside  the
external energy triples. The explicit decision between the two
existing possibilities of these $E \neq 0$ mergers will be
controlled by the detailed balance between the size of $B$ and $C$
-- for the weaker $C$ the pattern will resemble Figure
\ref{obr2b}.

The last possibility corresponds to the full dominance of the
constant $B$. This weakens the attraction between the central and
peripheral energy pairs. A characteristic illustration is offered
by Figure  \ref{obr4} where the choice of $A=B/2=C=1$ is shown to
lead to the spectrum characterized by a central-pair-attraction
dominance and by the related loss of quasi-Hermiticity at
$t^{(QH)}=0.5157267$. Our last Figure \ref{obr5} then shows how
the other, peripheral-pair-attraction dominance modifies the
previous result. With the choice of $A=3$, $B=5$ and $C=1$ we get
the triplet of pseudo-Hermiticity boundaries
$t^{(QH)}=0.5436890127$, $t^{(QH)}=0.4693964246$ and
$t^{(QH)}=0.4273046236$. The model loses its quasi-Hermiticity at
$t<t^{(QH)}=0.539764657$.

\section{The general chain model with $N=2J$ levels
 \label{generel} }

One could move on and construct various sample spectra,
numerically, at a number of the higher even dimensions $N=2J$. The
same family of the matrix models as mentioned in the previous
section would still suit our purpose. The three main user-friendly
features of all these models can be seen

\begin{itemize}

\item
 in the fact that they represent a generalization of eq.~(\ref{6IPTS}),

\item
in the nontrivial fact that the same parametrization  can be
used,
 \ben
 g_n=\sqrt{ g_n^{(max)} \left(1-\xi_n \right) }\,,
 \ \ \ \ \ \ \ \ \
 \xi_n = t+t^2+\ldots+t^{J-1}+G_n t^J\,,
 \ \ \ \ n = 1, 2, \ldots, J\,
 \een

\item
 in the fairly nontrivial recommendation that $g_n^{(max)}=n\,(N-n)$
(cf. \cite{maximal}).

\end{itemize}

 \noindent
The numerical experiments of the two preceding sections indicate
that the energy mergers in the spectra might admit a combinatorial
classification. In such a perspective, Figure \ref{obr1a} is still
trivial since for the mere two available energy levels at $J=1$
there exists just their single possible merger. Still, this
example enables us introduce a new convention that the two
energies will be subscripted by their respective unperturbed
integer values at $g_1=0$ (giving $E_{\pm 1}(t)$ at $J=1$, etc).

In this notation, the pair of the Figures \ref{obr2b} and
\ref{obr2c} with $J=2$ documents that among the four available
energies, there exist just the two topologically nonequivalent
mergers. In the first case the evolution in $-t$ connects $E_3$
with $E_1$ and $E_{-1}$ with $E_{-3}$, i.e., in an ordered
shorthand notation, $\{[-3,-1],[1,3]\}$. In the second case one
connects $E_3$ with $E_{-3}$ and $E_{1}$ with $E_{-1}$ and gets
the second symbol, $\{[-3,3],[-1,1]\}$.

In such an approach we ignore the transitional multiple mergers as
sampled in Figures \ref{obr2a} or \ref{obr3a}. Still, the
classification of the non-intersecting pairwise connections
between the $N=2J$ energy levels requires the knowledge of the
number $P^{(2J)}$ of nonequivalent connections among the levels at
every even dimension $2J$.  For its evaluation let us first order
the energies in a left-right symmetric string, or lattice,
$E_{-(2J-1)}$, $E_{-(2J-3)}$, $\ldots$, $E_{2J-3}$, $E_{2J-1}$.
This indicates that our pattern of connections must be left-right
symmetric in this visualization.

We already know that $P^{(2)}=1$ and $P^{(4)}=2$. At $J=3$ and
$N=6$ it is still easy to check that just the following three
different possibilities exist,
 \ben
 \{[-5,-3],[-1,1],[3,5]\}\,,
 \ \ \ \ \
 \{[-5,5],[-3,-1],[1,3]\}\,,
 \ \ \ \ \
 \{[-5,5],[-3,3],[-1,1]\}
 \,,
  \een
i.e., $P^{(6)}=3$. This enumeration results not only from the
required left-right symmetry but also from the necessary absence
of intersections between the individual energy-connection curves.

Elements of the sequence of the counts of the possibilities
$P^{(2J)}$ will be now generated in recurrent manner. Firstly, it
is obvious that the number $P^{(N)}$ of all the possible
arrangements of the mergers at a given $N$ always incorporates a
contribution from the subset of $P^{(N-2)}$ merging patterns which
all contain the longest possible merger $[-(2J-1),2J-1]$ of the
two outermost energies. Let us, therefore, abbreviate
$P^{(N)}-P^{(N-2)}=Q^{(N)}$ and, in the analysis of the remaining
$Q^{(2J)}$ options, let us distinguish between the even and odd
$J$.

In the former case with $J=2K$ none of the connections $[\pm
(2J-1),m]$ involving one of the outermost energies can ever cross
the center of the lattice. We have to fix ${\rm sign}\, m ={\rm
sign}\, (\pm (2J-1))\,\equiv\,\pm 1$. Thus, the absolute value of
the index $m$, skipping always a pair of the levels, will run over
the $K-$plet of the odd integers $2J-3=4K-3$, $2J-7=4(K-1)-3$,
$2J-11=4{K-2}-3$, $\ldots$, $4-3=1$. This means that we may
introduce an auxiliary symbol $P^{(0)}=1$ and evaluate,
recurrently, $Q^{(4K)}$ as a sum of $K$ terms at any $K$. This
rule forms our first recurrence relation,
 \ben
 P^{(4K)}-P^{(4K-2)}=
 \een
 \ben
 =P^{(2K-2)}\cdot P^{(0)}+
 P^{(2K-4)}\cdot P^{(4)}+P^{(2K-6)}\cdot P^{(8)}
 +\ldots+P^{(2)}\cdot P^{(4K-8)}
 +P^{(0)}\cdot P^{(4K-4)}\,.
 \een
In parallel, the choice of the odd $J=2L+1$ makes the
second-longest connections $[\pm (2J-1),\pm 3]$ a bit shorter,
leaving a two-point gap in the middle of the lattice.  This means
that the sequence of all the connections $[\pm (2J-1),\pm n]$
which involve the outermost energies will again possess $L$ terms.
We arrive at the second recurrence relation,
 \ben
 P^{(4L+2)}-P^{(4L)}=
 \een
 \ben
 =P^{(2L-2)}\cdot P^{(2)}+
 P^{(2L-4)}\cdot P^{(6)}+P^{(2L-6)}\cdot P^{(10)}
 +\ldots+P^{(2)}\cdot P^{(4L-6)}
 +P^{(0)}\cdot P^{(4L-2)}\,.
 \een
The latter two recurrences are mutually coupled. Their numerical
solution is straightforward, with a sample given in
Table~\ref{pexp2}. Some of its properties are really remarkable.
For example, empirically one finds out that the first eight (!)
elements of the sequence $P^{(2J)}$ coincide with certain binomial
coefficients. For the {\em next  eight} elements of this sequence,
moreover, one still finds another unexpected regularity in the
differences
 \ben
 R^{(K)}=\frac{1}{2}\,\left [P^{(4K)}
 -
 \left (
 \ba
 2K\\K
 \ea
 \right )
 \right ]\,,
 \ \ \ \ \ \ \ \
 S^{(K)}=\frac{1}{4}\,\left [P^{(4K+2)}
 -
 \left (
 \ba
 2K+1\\K
 \ea
 \right )
 \right ]\,
 \een
(cf. Table~\ref{pexp2}).

\begin{table}[t]
\caption{Multiplicities $P^{(N)}$ of the merging patterns}
\label{pexp2}
\begin{center}
\begin{tabular}{||c|ccccccccccc||}
\hline \hline
 $K$&0&1&2&3&4&5&6&7&8&9&
 \ldots\\
 \hline
 $P^{(4\,K)}$&
 1&2&6&20&68&234&808&2798&9700&
    33656&
    \ldots \\
  $P^{(4\,K+2)}$&
 1&3&10&35&122&426&1484&5167&
    17974&62498&
    \ldots \\
    \hline
 $R^{(K)}$&
 0&0&0&0&1&9&58&317&1585&
    7482&
    \ldots \\%
 $R^{(K)}-S^{(K)}$&
 0&0&0&0&0&0&0&0&1&12&
 \ldots \\
 \hline \hline
\end{tabular}
\end{center}
\end{table}

\section{Conclusions \label{katak} }

We may summarize that our detailed quantitative analysis of
schematic matrix models revealed interesting generic qualitative
features of the important phenomenological concept of quantum
instabilities.

Firstly we saw that the very possibility of the conditional,
parameter-controlled emergence of the quantum collapse is closely
bound to the manifestly non-Hermitian character of the underlying
Hamiltonians. Indeed, these operators only {\em rarely} enable us
to suppress the well known robust mathematical stability of the
spectra when they are chosen as manifestly Hermitian.

Of course, the removal of the Hermiticity (in the {\em narrow}
sense of the invariance with respect to the matrix transposition
and complex conjugation) {\em does not} lead to any conflict with
the postulates of Quantum Mechanics. On the contrary, it enables
us to make the models more flexible and more amenable to a direct
control of the  mechanism of the complexification of the
eigenvalues.

In a way related to the simplicity of our examples another key
merit of them can be seen in the {\em nontriviality} of the
related metric $\Theta \neq I$ which can (and does) vary with the
parameters. As long as the flexibility of the physics is directly
encoded in $\Theta$, one can conclude that the access to the onset
and/or breakdown of the observability should be mediated by the
selection of the ``decisive" parameters.

We did not mention many other merits of our models (like, e.g.,
the particular advantages of their ${\cal PT}-$symmetry, etc)
because the arguments in this direction may be found elsewhere
\cite{Carl}. For compensation, let us finally note that the
present outline of some properties of the quasi-Hermiticity
domains could be understood, in some sense, as the first steps
towards the formulation of a certain quantum analogue of the
Thom's theory of catastrophes \cite{Thom}.


\vspace{5mm}

\section*{Acknowledgement}

Correspondence and discussion partnership of Hendrik B. Geyer and
Frederick G. Scholtz are gratefully appreciated. Monetarily
supported by the GA\v{C}R grant Nr. 202/07/1307, by the M\v{S}MT
``Doppler Institute" project Nr. LC06002 and by the NPI
Institutional Research Plan AV0Z10480505.

\vspace{5mm}

\section*{Table captions}

\subsection*{Table  \ref{pexp2}.
Multiplicities $P^{(N)}$ of the merging patterns}

\vspace{5mm}

\section*{Figure captions}

\subsection*{Figure  \ref{obr1a}. Real parts of the energies in the
two-state model as functions of the parameter $t$.
 }

\subsection*{Figure  \ref{obr1b}. Imaginary parts of the energies in the
two-state model as functions of the parameter $t$.
 }

\subsection*{Figure  \ref{obr2a}. The four real $J=2$ energies at
$A=B=1$
 }

\subsection*{Figure  \ref{obr2b}. The four real $J=2$ energies at
$A/2=B=1$
 }

\subsection*{Figure  \ref{obr2c}. The four real $J=2$ energies at
$A=2B/3=1$
 }

\subsection*{Figure  \ref{obr2d}. The four real $J=2$ energies at
$A=B/5=1$
 }

\subsection*{Figure  \ref{obr3a}. Real parts of the energies in the
six-state model ($A=B=C=1$).
 }

\subsection*{Figure  \ref{obr3b}. Imaginary parts of the energies in the
six-state model ($A=B=C=1$).
 }

\subsection*{Figure  \ref{obr4}. Real parts of the energies in the
six-state model at $A=B/2=C=1$.
 }

\subsection*{Figure  \ref{obr5}. Real parts of the energies in the
six-state model at $A/3=B/5=C=1$.
 }

\end{document}